\documentclass[aps,prapplied,twocolumn,amsmath,amssymb, superscriptaddress,nofootinbib,reprint]{revtex4}
\usepackage{graphicx}
\usepackage{dcolumn}
\usepackage{bm}
\usepackage{hyperref}
\usepackage{epstopdf}
\usepackage{amsmath, amsthm}
\usepackage{xcolor}
\usepackage{multirow}

\usepackage[caption=false]{subfig}
\def \bea {\begin{eqnarray}}
\def \eea {\end{eqnarray}}
\def \be {\begin{equation}}
\def \ee {\end{equation}}
\def \asi{{\it a}-Si}
\def \asio2{{\it a}-SiO$_2$} 

\def \deg {${^\circ}$}

\begin{document}

\title{Realistic inversion of diffraction data for an amorphous solid: the case of amorphous silicon}

\author{Anup Pandey}
\affiliation{Department of Physics and Astronomy, Condensed Matter and Surface Science Program, Ohio University, Athens OH 45701, USA}

\author{Parthapratim Biswas}
\affiliation{Department of Physics and Astronomy, The University of Southern Mississippi, Hattiesburg MS 39406, USA}

\author{Bishal Bhattarai}
\affiliation{Department of Physics and Astronomy, Condensed Matter and Surface Science Program, Ohio University, Athens OH 45701, USA}

\author{D. A. Drabold}
\affiliation{Department of Physics and Astronomy, Nanoscale and Quantum Phenomena Institute, Ohio University, Athens OH 45701, USA}

\date{\today}

\begin{abstract}
We apply a new method ``force enhanced atomic refinement" (FEAR) to 
create a computer model of amorphous silicon ({\it a}-Si), based upon 
the highly precise X-ray diffraction experiments of Laaziri et 
al.~\cite{laaziri}. The logic underlying our calculation is to 
estimate the structure of a real sample {\it a}-Si using experimental data and chemical information included in a non-biased way, starting from random coordinates.  
The model is in close agreement with experiment {\it and} also sits at a suitable minimum energy according to density functional calculations. In agreement 
with experiments, we find a small concentration of coordination defects that we discuss, including their electronic consequences. The gap states in the FEAR
model are delocalized compared to a continuous random network model.  The method is more efficient and accurate, in the sense of fitting the diffraction data than conventional melt quench methods. We compute the vibrational density of states and the specific heat, and find that both compare favorably to experiments.

\end{abstract}

\pacs{61.43.-j, 61.43.Bn, 71.15.Pd, 71.23.-k}

\maketitle

\section{Introduction}

It has long been realized that the inversion of diffraction data -- extracting a structural model based upon the data at hand -- is a difficult problem of materials theory. It is worth noting that the success of inverting diffraction data for crystals has been one of the profound success stories of science, even revealing the structure of the Ribosome~\cite{venki}. The situation is different for non-crystalline materials. Evidence from Reverse Monte Carlo (RMC) studies~\cite{rmc_biswas,pusztai,rmc2,mcgreevy} show that the information inherent to pair-correlations alone is not adequate to produce a model with chemically realistic coordination and ordering. This is not really surprising, as the structure factor $S(Q)$ or pair-correlation function $g(r)$ (PCF) is a smooth one-dimensional 
function, and its information entropy~\cite{jaynes} is vastly higher (and information commensurately lower) than for a crystal, the latter PCF being a sequence of sharply localized functions. It seems clear that including chemical information, {\it in an unbiased mode}, should aid the structure determination substantially. Others have clearly described this challenge as the ``nanostructure problem"~\cite{billinge}, and noted the appeal of including an interatomic potential. We show here that such an approach is successful, by uniting the RMC code ``RMCProfile" and including chemistry in a self-consistent manner using density functional theory, but not by invoking {\it ad hoc} constraints. We have named this method ``Force Enhanced Atomic Refinement" (FEAR). In this paper, we focus on the classic and persistently vexing problem of amorphous silicon.  The details of the methods can be found elsewhere ~\cite{fear1,fear2}. The method is fast enough to make it easy to implement with {\it ab initio} interactions 
({\sc Siesta} here) and plane-wave DFT ({\sc Vasp}) as we used in ternary chalcogenide materials in 
Ref.\,~\cite{fear2}. 

The technological importance of {\asi} in microelectronics, thin-film transistors and  photo-voltaic (PV) applications~\cite{street} has led to many studies in recent decades~\cite{santos,pany,art,laaziri,pandey1,roorda}. In addition, the over-constrained network makes the structure of {\asi} difficult to model~\cite{Beeman-PRB_1985,www}. The only method that produces really satisfactory models for {\asi} is the Wooten-Weaire-Winer (WWW)~\cite{www} scheme, which is limited by unrealistic interactions and is also not a general technique. 

From a practical modeling perspective, the utilization of {\it a priori} information by constraining chemical order and preferred coordination has improved some of the most serious limitations of RMC~\cite{taf_dad}. Cliffe and co-workers imposed `uniformity' as a constraint in a refinement of atomistic-scale structures in their INVariant Environment Refinement Technique (INVERT)~\cite{goodwin}, and considerably extended their analysis by invoking `structural simplicity' as a guiding principle in modeling a-Si\cite{matt}. Recently another angle has been tried: including electronic {\it a priori} information in the form of an imposed band gap ~\cite{gap_sculpting, pssapaper}. These constraints are externally imposed and sensible though they might be, they introduce the investigators {\it bias} in the modeling. In other applications, more along the lines of "Materials by Design" the point is indeed to impose conditions that the model must obey -- and see if a physical realization of the desired properties may be realized. This is beyond the scope of the present paper which is focused on trying to best understand well explored specific samples of {\asi}. 

More in the spirit of our work, a hybrid reverse Monte Carlo (HRMC) incorporating experimental data and a penalty function scheme was introduced to find models of amorphous carbon in agreement with diffraction data also near a minimum of an empirical potential~\cite{opletal}. Gereben and Pusztai employed a similar approach of hybrid RMC with bonded and non-bonded forces to study liquid dimethyl trisulfide~\cite{rmc_pot}.  The first attempt to incorporate experimental information in a first-principle approach was experimentally constrained molecular relaxation ({\sc Ecmr})~\cite{ecmr,biswas_modelling}. {\sc Ecmr} merely alternated full relaxations of fitting pair-correlations (via RMC) and energy minimization. When this process converged (as it did for the case of glassy GeSe$_2$), an excellent model 
resulted~\cite{ecmr}. The problem was that this scheme often failed to converge.  We 
therefore amended {\sc Ecmr} and introduced {\it ab initio} force-enhanced atomic refinement ({\it ab initio} FEAR)~\cite{fear2}. In effect we alternate between partially fitting the RDF (or structure factor) using RMC and carrying out partial relaxations using {\it{ab initio}} interactions, as we explain in detail in References \onlinecite{fear1,fear2}. By carrying out the iteration in ``bite-sized" bits rather than iterated full relaxations as in the original {\sc Ecmr}, we find that the
method is robust, working for silver-doped chalcogenides with plane-wave DFT and 
for WWW {\asi} with {\sc Siesta} and also for forms of 
amorphous carbon~\cite{fear4}.

 We should clarify that in our previous work on {\asi}~\cite{fear2}  we used the WWW RDF as input ``experimental data", whereas in this work we have used high energy X-ray diffraction data from Laaziri et al.~\cite{laaziri}. WWW models are a fixture of the modeling community (a continuous random network of ideal four-fold coordination and involving up to 100,000 atoms~\cite{art, draboldurbach}, and represent an important benchmark that a new method must handle. It is reasonably interpreted as ``ideal" a-Si, with minimum strain. While the RDF of WWW and Lazirri~\cite{laaziri} are indeed fairly similar, there are key differences as noted by Roorda and coworkers~\cite{roorda}. Given the high quality and precision of the experiments, we have undertaken a FEAR inversion of their data in this paper. 
 
One key assumption that we forthrightly emphasize is that the dataset of Laaziri and coworkers may be represented by a small supercell model of silicon. This is obviously an approximation, as the material must surely include some voids and damaged regions from the ion bombardment procedure from which the material was made, and of course the X-ray diffraction includes these. While we think this is a reasonable approximation, it is clear that a very large scale simulation with thousands of atoms allowing for internal surfaces and other inhomogeneities would be desirable, possibly opening up the possibility of paracrystallites~\cite{treacy} and other longer length scale irregularities. It is not obvious whether the RDF by itself would provide information enough to open up voids. Such computations might be undertaken with transferable potentials devised from ``machine learning"\cite{machine}.
 
In our applications of {\sc FEAR}, we have always started with a {\it random} model, and even for a complex ternary~\cite{fear2} the method converges with satisfactory and chemically sensible results. In effect, chemical information is provided through the partial CG relaxations, and the method explores the configuration space rather well, thanks to the excellent RMCProfile code~\cite{rmcprofile}.  

The rest of this paper is organized as follows. In Section II, we summarize FEAR and describe the methodology for the current work.
In Section III, we present the results of FEAR for a 216-atom {\asi} model. The conclusions are given in Section IV.
\begin{figure}
\begin{center}
\includegraphics[width=2.8 in]{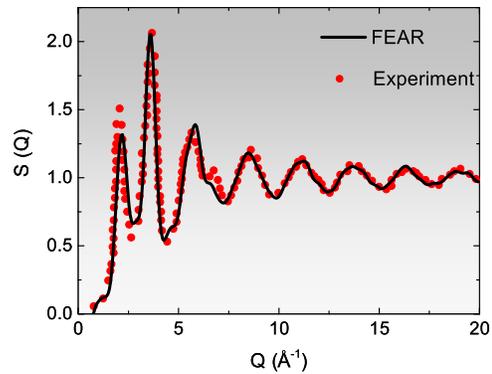}%
\caption{(Color online) 
Comparison of the simulated X-ray static structure factor (black) 
from FEAR with the experimental diffraction data (red circle) 
from Ref.\,\onlinecite{laaziri}. A 216-atom model is used to 
produced the simulated structure factor. 
}
\label{sqall}
\end{center}
\end{figure}
\begin{figure}
\begin{center}
\includegraphics[width=2.8 in]{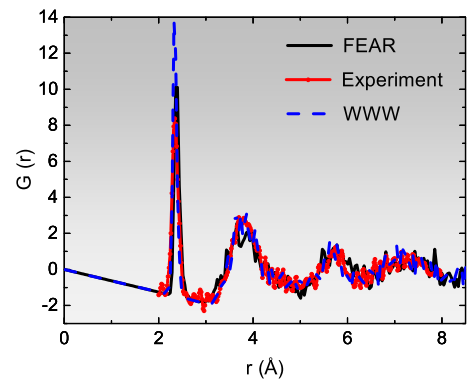}%
 
\caption{(Color online)
The reduced pair-correlation function of 
{\asi} obtained from a 216-atom model using FEAR 
(black) and WWW (blue) methods. The experimental data (red)
shown above are the Fourier transform of the 
high-energy X-ray diffraction data from 
Ref.\,\onlinecite{laaziri}.
}
\label{Grall}
\end{center}
\end{figure}
\begin{figure}
\begin{center}
\includegraphics[width=2.8 in]{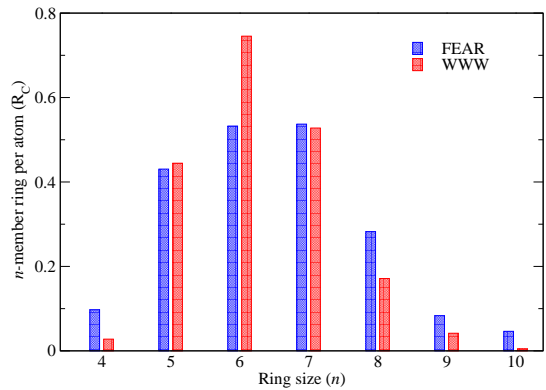}%
 
\caption{(Color online)
The number of \emph{n}-fold ring per atom (R$_{C}$) for the FEAR model (blue) compared to the WWW model of same size. 
}
\label{rings}
\end{center}
\end{figure}
\section{Methodology}

More details about FEAR can be found elsewhere~\cite{fear1,fear2}. To summarize, in FEAR, 
a {\em random} starting configuration is subjected to partial RMC refinement followed by partial conjugate gradient (CG) relaxation according to a chemically realistic (say DFT interaction). The two steps are repeated until both the structure and energy converge~\cite{fear1,fear2}. In this work, we have carried out RMC for 500 accepted moves followed by 5 CG relaxations steps (we have tried other recipes such as 1000 and 10 moves, respectively, with similar results). This process is then repeated until convergence (namely finding coordinates both matching diffraction data and being at a minimum of a DFT total energy). The RMC algorithm (in our case RMCProfile~\cite{rmcprofile}) is used to invert the experimental data. We have so far only used diffraction data, though EXAFS and NMR are also natural datasets to attempt, and in principle multiple experimental datasets might be jointly fit while the CG relaxations enforce chemistry in the material. We employ a local-orbital basis DFT code ({\sc Siesta})~\cite{siesta} using the local density approximation (LDA).  The cubic box edge length is 16.281 {\AA} which corresponds with the experimental density of 2.33 gm/cm$^{3}$ (which, in the spirit of full reporting should be understood to be another {\it assumption}). 

\begin{figure}
\begin{center}
\includegraphics[width=2.8 in]{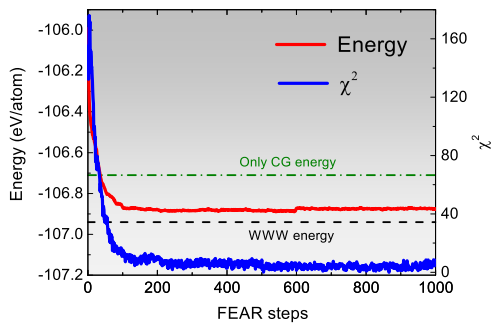}%
 \caption{(Color online) Total energy per atom and $\chi^{2}$ versus 
FEAR steps for a 216-atom {\asi} model. The green and black broken 
lines represent the energy per atom for the CG-only and WWW 
model, respectively.}
\label{egy_all}
\end{center}
\end{figure}

\section{Results and discussion} 

In this section, we present results for {\asi} obtained from 
FEAR. Since the FEAR method 
essentially consists of incorporating the 
pair-correlation data via reverse monte carlo 
simulations (RMC),  followed by {\it ab initio} 
total-energy relaxations using the conjugate-gradient 
(CG) method, we also include the results from the 
CG-only model (e.g. from the initial random state (to evaluate the performance of
FEAR with the CG method as a function of 
simulation time or steps.  In particular, we address the structure factor $S(Q)$, bond-angle 
distribution $P(\theta)$, electronic density of states (EDOS), 
the vibrational density of states (VDOS) and the vibrational specific heat of the 
FEAR models of {\asi}. To examine the convergence of the 
method with respect to the total energy and the evolution
of structure, we take a close look at the variation of the
average coordination number and optical gap as a function of FEAR steps.

Figure \ref{sqall} shows the structure factor of {\asi} for 
the model configurations obtained from the FEAR along with the (annealed sample) structure-factor data 
of {\asi} reported by Lazirri et al.~\cite{laaziri}. Fitting was 
carried out in Q space. It is apparent that, while the CG-only 
model shows a consistent deviation from the experimental data, 
particularly at high $Q$ values, the structure factor from 
the FEAR model compares very well with the experimental data. The only exceptions are a 
minor deviation of $S(Q)$ near $Q$=2.5 {\AA} 
and 7 {\AA}.  
A comparison of the $S(Q)$ data from the FEAR and CG-only models 
suggests that the former is superior to the latter 
as far as the two-body correlations of the models 
are concerned even though both the systems have been 
treated with identical {\it ab initio} interactions. 
This observation is also reflected on Fig.\,\ref{Grall}, where 
the reduced radial distribution function, $G(r) = 4 \pi r n_0 (g(r)-1)$,
obtained from FEAR, WWW, and X-ray diffraction 
experiments are plotted. 

Since the pair-correlation data or structure factors of a model 
cannot determine a three-dimensional amorphous structure 
uniquely, it is necessary to examine the models further by going beyond two-body correlation 
functions. To this end, we have calculated the bond-angle distribution 
$P(\theta)$, and compared it with the results obtained from 
WWW, CG-only and the width of the transverse optical (TO) peak of the 
Raman spectrum of {\asi}. Following Beeman et al.\,\cite{Beeman-PRB_1985}, 
we have assumed that the half-width at half-maximum (HWHM) of the Raman 
TO peak of {\asi} is related to the average width of the bond-angle 
distribution. Since a typical value of the width of the Raman TO peak 
in {\asi} ranges from 33 to 50 cm$^{-1}$, this approximately 
translates into a value of 9-13\deg for the average bond-angle 
deviation. This value is not far from with the RMS angular 
deviation (HWHM) of 15.6\deg from the FEAR model.  It is 
noteworthy that the FEAR model is statistically free 
of very small ($\le$ 60\deg) or large ($\ge$ 160\deg) angles, 
and that the bond-angle distribution closely matches with the 
same from the WWW model. In contrast, a considerable number 
of small and large angles, below 60{\deg} and above 160{\deg}, 
respectively, have appeared in the bond-angle distribution of 
the CG-only model and in the RMC-only model\cite{fear2}. 
Thus, the FEAR method not only produces correct two-body 
correlations between atoms, but also a better {\it reduced} 
three-body correlations by judicious use of the input 
experimental data and the local chemical information of {\asi} 
provided from the {\it ab initio} total-energy functional from 
{\sc Siesta} within the CG loop of the refinement process.  
We have compared the ring statistics for the FEAR model to that of the WWW model in Fig.~\ref{rings}. Three-member rings are absent in both FEAR and WWW model which is consistent with the absence of unphysical Si triangles in good quality models. The only notable difference between the WWW and FEAR model is the existence of fewer 6-member rings in the FEAR model.

In Table \ref{tab1}, we have listed the characteristic structural properties of 
the models along with the total energy per atom obtained from the 
density-functional code {\sc Siesta}. The FEAR model has 96\% 
four-fold coordination, with equal (2\% fractions) of 3-fold and 5-fold Si. This is equal to the melt-quench model using environment-dependent interaction potential (96\%)~\cite{md_edip} and better than models obtained from other techniques~\cite{rmc_biswas,goodwin,md_tersoff}.
The average coordination number of our model is 4 which deviates from that of the experimental annealed sample (3.88)~\cite{laaziri}. For comparison we have presented average coordination for various models in Table \ref{tab1}. It appears that the models having fewer coordination defects have higher average coordination then the experimentally reported value. 

\begin{figure}
\begin{center}
\includegraphics[width=0.5 \textwidth]{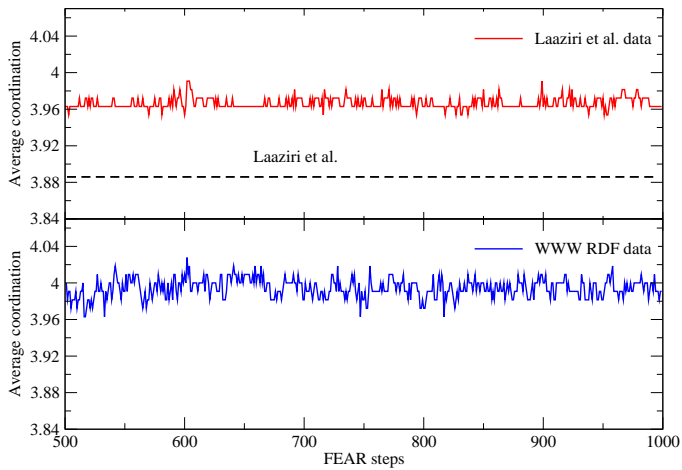}
\caption{
(Color online) Variation of the average coordination number 
for the final 500 steps of FEAR using two different input RDF 
data. The upper panel is for high-energy X-ray diffraction 
data from Laaziri {\it et al.}~\cite{laaziri} and the lower 
panel is for the WWW radial distribution function (RDF) 
as an input data~\cite{fear2}. The broken horizontal 
line,  in the upper panel, represents the average 
coordination number, 3.88,  reported by Laaziri 
{\it et. al.}~\cite{laaziri}
}
\label{coordination}
\end{center}
\end{figure}

The variation of the total energy (E) and $\chi^2$ FEAR proceeds is indicated in Fig.\ref{egy_all}.  
Figure \ref{egy_all} suggests that the initial structure formation 
takes place very rapidly in the first few hundred steps with the 
simultaneous decrease of $E$ and $\chi^2$. We then reach a period of ``saturation" in which
there are tiny fluctuations in the energy and $\chi^2$. We have reported a particular ``snapshot" of a conformation, and discuss it above. However, the many conformations in the 
saturated part of the computation are equally meaningful. Fortunately they do not fluctuate much, reflecting the fact that the combination of experimental data and chemistry converge
to a well-defined collection of configurations. We track the fluctuations in mean coordination in Fig. \ref{coordination}, excised from the last 500 steps of FEAR. For convenience we also show the results for a simulation with the WWW RDF, as we report in Ref. \onlinecite{fear2}. Using the RDF of WWW as input data forces the network to have fewer defects compared to the real experimental data.  
FEAR for the experimental sample fluctuates around 3.96, whereas the WWW fluctuates around 3.99.

\begin{figure}
\begin{center}
\includegraphics[width=0.5 \textwidth]{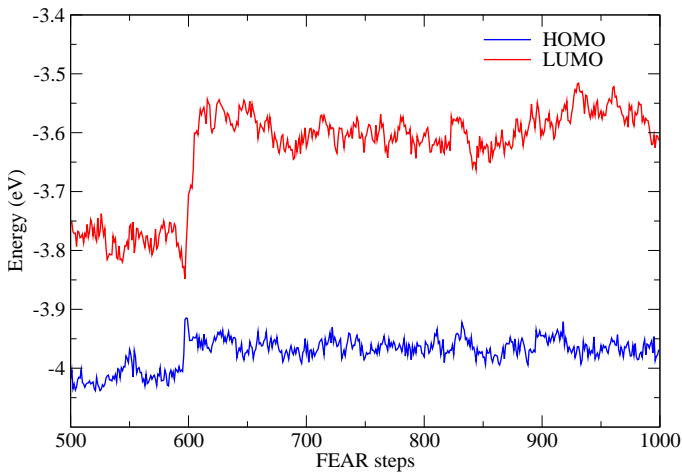}
\caption{
(Color online) Variation of the highest occupied molecular orbital (HOMO)level and the lowest unoccupied molecular orbital (LUMO) level for the final 500 steps of FEAR.
Note the annihilation of an electronic (gap state)  defect near 600 steps.
}
\label{homolumo}
\end{center}
\end{figure}

In Fig. \ref{homolumo}, we also track the fluctuations in the optical gap for the last 500 steps of FEAR, as crudely estimated as the energy splitting between the LUMO and HOMO levels. It is of considerable interest that for the last 500 FEAR steps, there is a substantial variation in the electronic density of states near the Fermi level even though the FEAR process had already reached a ``steady state" value for $\chi^2$ and the total energy (compare Fig. \ref{egy_all}). Observe too that while the HOMO level is fairly stationary, the LUMO meanders with relative impunity no doubt because it
does not contribute to the total energy, being above the Fermi level. Thus, we see that FEAR effectively generates an ensemble of candidate structural models for a-Si which are essentially indistinguishable according to $\chi^2$ and energy. Nevertheless, this affords another opportunity to use {\it a priori} information -- we should select one of these models with the gap most like the experimental sample. To our knowledge, the electronic density of states is not well characterized for the sample, but if it was it would be natural to use it as an additional criterion to select the most experimentally realistic FEAR model. In effect if we had electronic information it would break the ``structural degeneracy", emphasizing the information-based nature of our approach.

It is evident from Fig.\,\ref{egy_all}
that the FEAR model has a lower energy than its CG-only 
counterpart.  Table \ref{tab1} lists the total-energy per atom 
w.r.t the energy of the WWW model, which is set at 0.0 eV for 
convenience.  The energy for the FEAR model is found to be 
0.06 eV/atom, which is approximately 50\% lower than 
the CG-only model with a total-energy of 0.09 eV/atom.  
This is a reasonable number compared to other published 
work~\cite{dadpssrrl}.

\begin{table}
\caption{ 
Total energy and key structural properties of {\it a}-Si models. 
The energy per atom is expressed with reference to the energy 
of the WWW model. 
}
 \begin{center}
\begin{tabular}[b]{|p {1.8cm}|p {1.5cm}|p {1.2cm}|p {1.2cm}|p {1.2cm}|}

  \hline
     &RMC & Only CG & FEAR & WWW \\ \hline
    4-fold Si (\%) & 27 & 75 & 96 &100 \\ \hline
    3-fold Si (\%) & 15 & 21 & 2 &0\\ \hline
    5-fold Si (\%) & 25 & 3 & 2 &0 \\ \hline
     Energy (eV/atom) & 3.84  &0.09 & 0.06 &0.00\\ \hline
     Average bond angle (RMS deviation)  & 101.57{\deg} (31.12\deg)  & 107.31{\deg} (20.42\deg)& 108.52{\deg} (15.59\deg) &108.97{\deg} (11.93\deg)\\ \hline
     Average coordination number & 4.27  & 3.83 & 4.00 &4.00\\ \hline
      \hline
\end{tabular}
\label{tab1}
\end{center}
\end{table}

\begin{figure}
\begin{center}
\includegraphics[width=2.8 in]{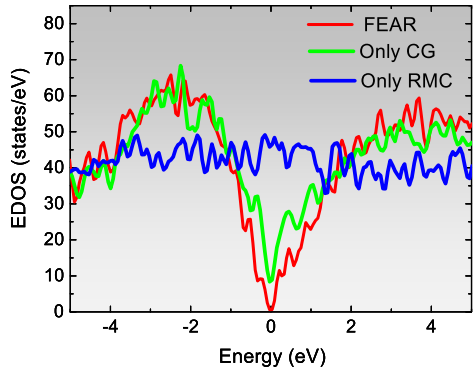}%
\caption{(Color online) Electronic density of states (EDOS) 
of {\asi} obtained from FEAR (red), 
CG-only (green) and pure RMC (blue) models. The Fermi levels 
are located at 0 eV.
}
\label{dosall}
\end{center}
\end{figure}

\begin{figure}
\begin{center}
\includegraphics[width=2.8 in]{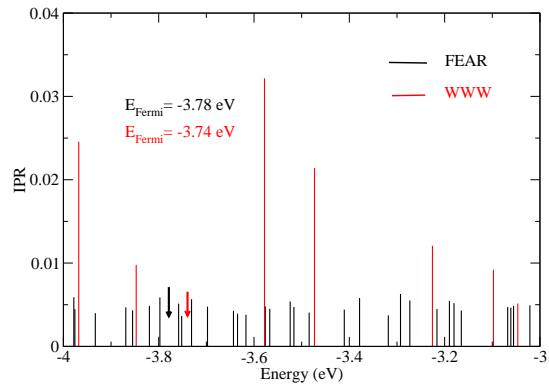}%
\caption{(Color online) Inverse participation ratio 
(IPR) of 216-atom {\asi} model for FEAR (black) and RMC (red) models near the gap. Fermi levels are shown by arrows of respective colors.}
\label{iprall}
\end{center}
\end{figure}

The electronic density of states (EDOS) of {\asi} 
obtained from the FEAR, CG-only and RMC models 
are shown in Fig.\ref{dosall}. For the 216-atom FEAR model, the quality of EDOS is significantly improved compared to that of CG relaxed model and the EDOS of the RMC model which is featureless. The significant number of defects states clutters the gap in FEAR, which is a prediction in this case, since the EDOS has not to our knowledge been measured for the sample we are studying.  Electronic localization is studied using the inverse participation ratio (IPR)~\cite{elec_ph} which is shown in Fig.\ref{iprall}. Banding among the states in the gap leads to an expected delocalization\cite{jjdadprl}. 

\begin{figure}
\begin{center}
\includegraphics[width=0.5 \textwidth]{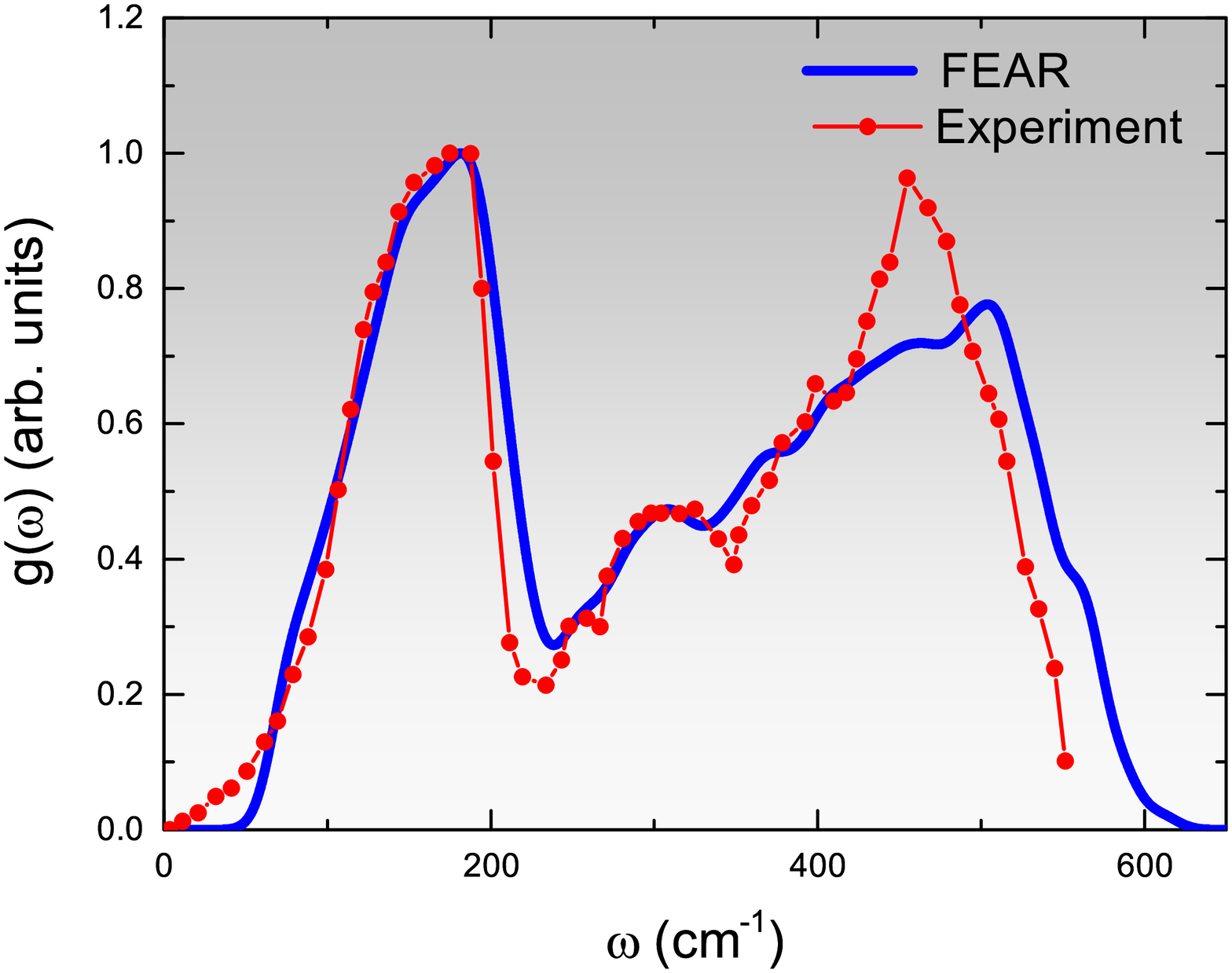}
\caption{
Vibrational density of states of {\asi}, $g(\omega)$,  from 
a 216-atom FEAR model (blue). The experimental vdos (red) 
obtained from Kamitakahara et al.~\cite{kamitakahara} 
}
\label{vdos216}
\end{center}
\end{figure}
\begin{figure}
\begin{center}
\includegraphics[width=0.4 \textwidth]{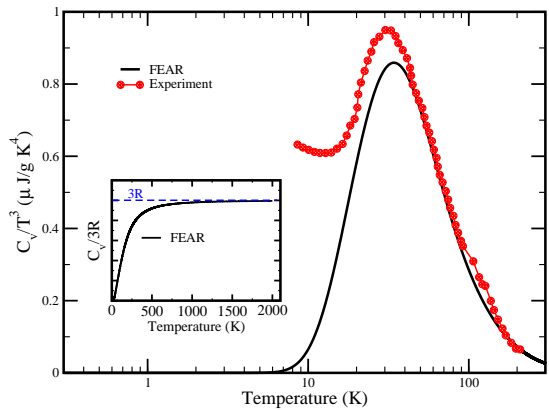}
\caption{
The specific heat capacity (C$_{V}$/T$^{3}$) for 216-atom {\asi} FEAR model (black) compared to the experiment~\cite{sp_heat_prl}. The inset shows the "Dulong-Petit" limit at higher temperature.
}
\label{specific_heat}
\end{center}
\end{figure}

The vibrational density of states (VDOS) is computed by estimating the force constant matrix, from finite-difference calculations resulting from perturbing the atoms of a well relaxed 216-atom FEAR model by 0.02 {\AA} in six directions ($\pm$ x-, $\pm$ y- and $\pm$ z- axis) and calculating the forces in all the remaining atoms for each perturbed configurations. The eigenvalues and eigenmodes are obtained by diagonalizing the dynamical matrix. The details can be found in a recent work of Bhattarai and Drabold ~\cite{bhattarai}. The VDOS for 216-atom model FEAR model is shown in Fig.\ref{vdos216}. The calculated vdos is in rather good agreement with the experimental vdos obtained from inelastic neutron scattering~\cite{kamitakahara}. The exception, probably a shortcoming of our Hamiltonian is a shift in the high frequency optical tail by $\sim$35 cm$^{-1}$. This observation 
is consistent with the other empirical and {\it ab initio} molecular-dynamics simulations~\cite{serge1,serge2}.

The specific heat in the harmonic approximation is easily obtained from the density of states, $g(\omega)$.
We note that wavelengths larger than our supercell size are not included in the obtained VDOS. We compute the specific heat $C_v(T)$ from the relation\cite{Maradudin1}$^,$\cite{Nakhmanson1},

\begin{equation}
 C(T)= 3R\int_0^{E_{max}}\Bigg(\frac{E}{k_{B}T}\Bigg)^2 \frac{e^{E/k_BT}}{\Big(e^{Ek_BT}-1\Big)^2}g(E)dE
\end{equation}

Here, the VDOS ($g(E)$) is normalized to unity.\\

In Fig.\ref{specific_heat}, we  see that $C_V(T)$ for FEAR  model is in a good agreement with experiment for $T>40K$~\cite{sp_heat_prl}.
The inset in Fig. \ref{specific_heat} indicates the ``3R'' (Dulong-Petit) limit at high temperature. This is an additional indication that the FEAR model
is reproducing features of a-Si beside those ``built in" (from the experimental data), and is also an indication of consistency between these very different
physical observables.

\section{Conclusions}

In this paper, we have studied {\asi} using a new approach FEAR. For the first time the experimental structure factor of {\asi}~\cite{laaziri} has been employed in FEAR along with {\it ab initio} interactions to generate a homogeneous model consistent with the data and at a plausible energy minimum according to reliable interatomic interactions. FEAR retains the simplicity and logic of RMC and successfully augments it with total-energy functional and forces to generate structures that are energetically stable, even exhibiting a satisfactory VDOS. The method can also be viewed as a new way to undertake first principles modeling of materials, when structural experiments are available.  

By using an entirely information-based approach, educated by chemistry through the CG sub loops, we find highly plausible models derived from experimental data with interesting similarities and differences with continuous random network models. Following this logic, the best that we can hope to achieve is a structural model jointly agreeing with all experiments, but critically, augmented with chemical information in an unbiased mode as we offer here.

\section{Acknowledgement}
We thank the US NSF under grants DMR 150683, 1507166 and 1507670 for supporting 
this work, and the Ohio Supercomputer Center for computer time. We acknowledge the financial support of the Condensed Matter and Surface Science program of Ohio University. We particularly thank Prof. Sjoerd Roorda for sharing his diffraction data, which formed the basis for this paper. We further thank Prof. Normand Mousseau for helpful suggestions.

\end{document}